\documentclass[%
reprint, nobalancelastpage,
superscriptaddress,
groupedaddress,
 amsmath,amssymb,
 aps,
 prx
linenumbers
]{revtex4-2}
\usepackage{comment}
\usepackage{braket}
\usepackage{graphicx}
\usepackage{dcolumn}
\usepackage{bm}
\usepackage{siunitx}
\usepackage{xcolor} 
\usepackage{tabularx}
\usepackage{dsfont}					
\usepackage{mathrsfs}				
\usepackage{float}
\usepackage{hyperref}

\newcommand{\be}{\begin{equation}}
\newcommand{\ee}{\end{equation}}
\newcommand{\bea}{\begin{eqnarray}}
\newcommand{\eea}{\end{eqnarray}}

\renewcommand{\vec}[1]{{\bf #1}}

\newcommand{\addAA}[1]{\textcolor{black}{#1}}

\usepackage{hyperref}
\hypersetup{colorlinks, 
	linkcolor={blue!75!black!80!yellow},
	citecolor={blue!75!black!80!yellow}, 
	urlcolor={blue!75!black!80!yellow}
	}

\usepackage[normalem]{ulem}

\usepackage{lineno}
\begin{document}
\title{Photo-induced superconducting diode effect via chiral cavity modes}
\author{Arpit Arora}
\email{arpit22@ucla.edu}
\affiliation{Division of Physical Sciences, College of Letters and Science, University of California, Los Angeles (UCLA), Los Angeles, CA, USA}
\affiliation{Department of Electrical and Computer Engineering, UCLA, Los Angeles, CA, USA}
\author{Prineha Narang}
\email{prineha@ucla.edu}
\affiliation{Division of Physical Sciences, College of Letters and Science, University of California, Los Angeles (UCLA), Los Angeles, CA, USA}
\affiliation{Department of Electrical and Computer Engineering, UCLA, Los Angeles, CA, USA}

\begin{abstract}
Time reversal symmetry breaking is an important facet of controlling nonreciprocal responses. Here, we propose a method of photo-control over superconducting diode-like nonreciprocities, where time reversal symmetry breaking is achieved via photon exchange with chiral cavity modes. We reveal the origin of the nonreciprocal superconducting response as the embedding of chirality in a many-body ground state through photon induced orbital magnetization. With twisted bilayer graphene (TBG) as an example, we demonstrate the general principles of photo-control of diode responses, which are valid for a wide range of superconductors and cavity designs. The cavity control of superconducting nonreciprocities, particularly in the microwave regime, offers a non-invasive means of exploring new functionalities in quantum circuits with ultrafast switching and on-chip integration. This control method can serve as an important contribution to the toolbox for nonreciprocal models in circuit quantum electrodynamics, primed to be harnessed for scalable and modular quantum devices.  
\end{abstract}
\maketitle
Photonic control of correlated matter offers a powerful route to advancing and manipulating the functionality of quantum materials~\cite{bloch2022strongly,hubener2024quantum}. The developments at the synergies of photonic and electronic matter have raised a fundamental question in the context of superconductivity: can photons serve as an active knob for the superconducting ground state? Many interesting phenomena such as photo-enhanced~\cite{Wyatt.1966,Dayem.1967,beck2013transient,Curtis.2019,lu2024cavity,arora2024quantum,kozin2025cavity,dhillon2026microwave} and photo-induced superconductivity~\cite{schlawin2019cavity,sheikhan2019cavity,andolina2024amperean,keren2026cavity} have emerged in an attempt to answer this question but the link between photonic control and superconducting nonreciprocities is still missing.

\begin{figure*}
    \centering
    \includegraphics[scale=0.43]{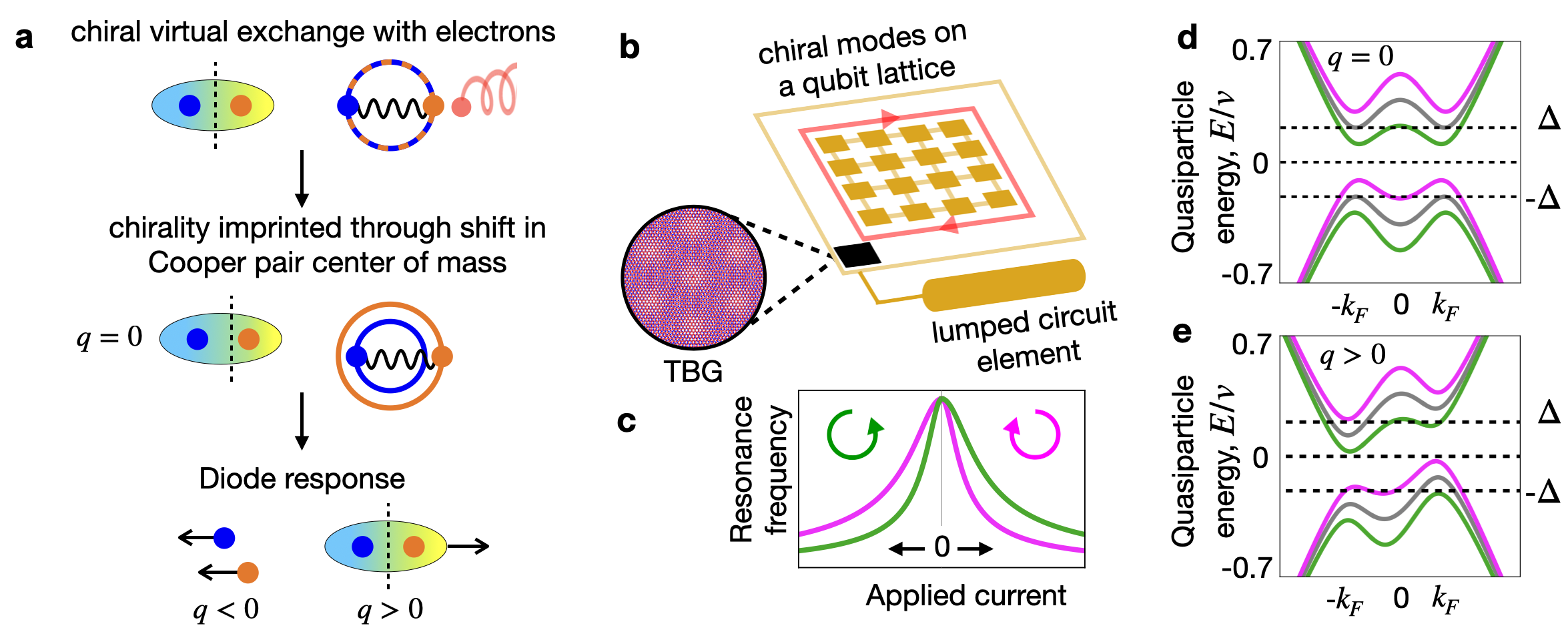}
    \caption{Cavity control of superconducting-diode like nonlinearities where $\mathcal{T}$-breaking is achieved via coupling of electronic system to chiral modes. (a) The chiral photon exchange with electrons shifts the Cooper pair center of mass (shown as a dashed vertical line) imprinting the chirality of photons onto the condensate. This leads to breaking of Cooper pairs dependent on direction of current, manifesting in a diode response. (b) Chiral modes on a qubit lattice coupled to superconducting TBG. (c) Schematic of a chirality dependent inductive response on a lumped circuit. (d,e) Chiral photons generate finite Cooper pair momenta which leads to chirality and thus to valley dependent alterations of the supeconducting gap; $\delta_c\neq 0,\chi=\pm$ (magenta, green) and $\delta_c = 0$ (gray).}
    \label{fig1}
\end{figure*}

Here we reveal a new paradigm for nonreciprocal superconductivity: \underline{C}hiral c\underline{A}vity co\underline{N}trol of superconducting \underline{D}iode-\underline{L}ike nonlin\underline{E}aritie\underline{S} (CANDLES). This provides a universal framework for photo-induced nonreciprocal superconducting responses where chiral photons coupled to Bloch electrons act as a source of orbital magnetic moments for Cooper pairs, imprinting handedness directly onto the condensate. The virtual photon exchange manifests as a chirality displacement of the Cooper pair center of mass, see Fig.~\ref{fig1}a, leading to an intrinsic asymmetry in stiffness and its excitation spectrum, giving a diode-like supercurrent without a magnetic field or engineered junction asymmetry.

Given the extensive use of superconductors in modern quantum architectures, superconducting diode-like nonlinearities have the potential to revolutionize quantum technology and open new frontiers in superconducting electronics~\cite{braginski2019superconductor,wendin2017quantum,blais2021circuit,schegolev2021superconducting}. In quantum hardware, such nonreciprocal elements can enable on-chip isolation and directional signal routing, protect qubit readout, and provide a low-loss alternative to centimeter-scale ferrite-based circulators, with clear potential for scalable modular quantum circuits~\cite{kannan2023demand,almanakly2025deterministic,barzanjeh2025nonreciprocity,dirnegger2025nonreciprocal}. While a number of underlying microscopic mechanisms have produced superconducting diode-like nonlinearities~\cite{ando2020observation,miyasaka2021observation,kawarazaki2022magnetic,wu2022nonreciprocal,nadeem2023superconducting,de2021gate,diez2023symmetry,lin2022zero,sundaresh2023diamagnetic,banerjee2024enhanced,huang2024superconducting}, overall, these diodes are classified as magnetoelectric phenomena, requiring simultaneous breaking of inversion ($\mathcal{I}$) and time reversal ($\mathcal{T}$) symmetries. In a typical setup, $\mathcal{T}$-breaking is controlled via an external magnetic field, which unfortunately also serves as a noise source in highly sensitive quantum circuits~\cite{blais2021circuit}.
Therefore, it is necessary to find non-invasive ways of controlling $\mathcal{T}$-breaking that perform better with quantum circuits. CANDLES enables an optimal avenue for such functionalities which can be realized 
with circularly polarized microwaves~\cite{mrozek2015circularly,yaroshenko2020circularly}, split-ring resonators~\cite{dembowski2003observation,plum2015chiral,peng2016chiral} and qubit lattices~\cite{owens2022chiral,du2024probing,labarca2024toolbox,levy2024passive,rosen2024synthetic}, e.g., see Fig.~\ref{fig1}b.

In this work, we show the principles of CANDLES through modified superconducting ground states under chiral photon exchange which are applicable to microwave, terahertz, and optical cavities.
We then demonstrate CANDLES in a twisted bilayer graphene (TBG) heterostructure interacting with chiral microwave modes. This leads to the formation of $\mathcal{T}$-breaking chirality-valley locked states manifesting in topological valley polarization, in contrast to previously studied spontaneous ferromagnetic valley polarization revealed through magnetic field-training~\cite{banerjee2024enhanced,scammell2022theory}. We also propose the measurement of diode responses with a lumped circuit element, see Fig.~\ref{fig1}c, using state-of-the-art hybrid circuit quantum electrodynamics devices~\cite{bottcher2024circuit,tanaka2024kinetic,banerjee2024superfluid,kreidel2024measuring} with estimates of the frequency resolution of such measurements for our results. 

\textit{$\mathcal{T}$-breaking in superconducting ground states: }
To lay out the fundamental understanding of CANDLES, we begin with a Bloch Hamiltonian, $H^0_\vec k$, interacting with chiral cavity modes, $\hat{\vec A} = \sum_{\lambda} A_0^\lambda(\vec e_{\lambda} a + \vec e_\lambda^*a^\dag)$. The chirality of these modes is given by $\vec e_{\lambda \in\{R,L\}} = (1,\chi i)/\sqrt{2}$ where $\chi=\pm$ for right ($R$) and left ($L$) handed modes, respectively. Importantly, the mode amplitude $A_0^\lambda = \sqrt{\hbar/2\varepsilon_0 V\omega_c^\lambda}$ is determined by cavity design in terms of mode confinement volume, $V$, and characteristic frequency $\omega_c^\lambda$ of mode $\lambda$; $\varepsilon_0$ is free space susceptibility.

The cavity field modifies the electronic system via minimal coupling, $H_0(\vec k - e\hat{\vec A}/\hbar)$. Focusing on microwaves, we consider the weak coupling limit and write the effective Hamiltonian for the Bloch system as $H_\vec k = H^0_{\vec k} + H^{1}_{\chi,\vec k} $. The effects of the chiral modes are captured in
\begin{equation}
\label{eq:intHamiltonian}
    H^1_{\chi, \vec k} = \chi g(A_0, n) C_\vec k
\end{equation}
where we assume complete chirality of modes. We define $g(A_0,n) = (1+2n)(e A_0)^2\omega_c/(2\hbar \Xi^2)$ which captures the effects of cavity parameters: $A_0$, and mode population, $n$ in Fock state $\langle a^\dagger a\rangle = n$, see Appendix A for details. Additionally, for microwave modes we have assumed $\omega_c \ll \Xi$ where $\Xi$ is the energy scale set by the lowest virtual transition in the electronic bands. A general expression for the interaction Hamiltonian is presented in Appendix A. Note that for optical/THz~\cite{dag2024engineering,jiang2024engineering,tay2025terahertz}, the interaction Hamiltonian is obtained by $\Xi\rightarrow \omega_c$ in Eq.~(\ref{eq:intHamiltonian}). 

Importantly, $C_\vec k = i[\hat{v}_\vec k^x,\hat{v}_\vec k^y]$ where $\hat{v}_{\vec k}^{x,y} = \partial_{k_{x,y}}H^0_\vec k$ captures the magnetic moment induced by chiral modes. Interestingly, $C_\vec k/\Xi^2$ represents the peak of Berry curvature in $\vec k$-space. $C_\vec k$ is odd under time reversal, i.e., $\hat{\mathcal{T}}C_\vec k = -C_{-\vec k}$ explicitly breaks $\mathcal{T}$. Note that for Berry curvature, $C_\vec k \neq 0$ only for multiple degrees of freedom (e.g., spin, valley, sublattice), underlining its close relation to topological responses such as the anomalous Hall effect, as emphasized in earlier works~\cite{mciver2020light,wang2019cavity,masuki2023berry,dag2024engineering,jiang2024engineering}. We advance this previous understanding to show how $C_{\vec k}$ directly enters the superconducting ground-state energy and current, making CANDLES a key principle of magnet-free, switchable superconducting diodes.

In the superconducting phase, we describe the system with a Nambu basis
\begin{equation}
\label{eq:bdgHam}
    \mathcal{H}_{\rm BdG}(\vec k,\vec q) = \begin{pmatrix}
        H_{\vec k+\vec q/2}-\mu & \Delta_\vec q \\
        \Delta_\vec q^\dagger & -H^T_{-\vec k + \vec q/2} + \mu
    \end{pmatrix}
\end{equation}
where $\mu$ is the chemical potential and $\vec q$ is the Cooper pair momentum. At zero temperature, the supercurrent is obtained as $\vec J_{\vec q} = \partial_{\vec q}E_{\rm GS}^{\vec q}$ through the superconducting ground state energy $E_{\rm GS}^\vec q = -\frac{1}{2}\sum_{\vec k,\alpha}E_{\vec k,\vec q,\alpha} + |\Delta_\vec q|^2/\mathcal{U}$ for pairing potential $\mathcal{U}$; $E_{\vec k,\vec q,\alpha}$ is obtained by diagonalizing $\mathcal{H}_{\rm BdG}$. The superconducting ground state is modified by chiral modes as $E_{\rm GS}\approx E_{\rm GS,0} -\frac{1}{2} \sum_{\vec k,\alpha}\chi g\langle \mathcal{M}\rangle_{\vec k,\vec q}^{\alpha\alpha}$ where $ \langle \mathcal{M}\rangle_{\vec k, \vec q}^{\alpha\beta} = \langle \Psi_{\vec k,\vec q,\alpha}^0|\mathcal{M}_{\vec k, \vec q} |\Psi_{\vec k,\vec q,\beta}^0\rangle$ with
\begin{equation}
\label{eq:GSenergy}
    \mathcal{M}_{\vec k, \vec q} = \begin{pmatrix}
        C_{\vec k + \vec q/2} & 0 \\
        0 & -C^T_{-\vec k + \vec q/2}
    \end{pmatrix}
\end{equation}
being the effective many-body magnetic moment operator obtained through generalization of $C_\vec k$ in Nambu space; $|\Psi_{\vec k,\vec q,\alpha}^0\rangle$ is calculated without the cavity. $\hat{\mathcal{T}}C_\vec k = -C_{-\vec k}$ enforces $\hat{\mathcal{T}}\langle\mathcal{M}\rangle_{\vec k, \vec q}^{\alpha\beta} = -\langle\mathcal{M}\rangle_{\vec k, -\vec q}^{\beta\alpha}$ in the presence of $\mathcal{I}$, thus requiring $\mathcal{T}$ and $\mathcal{I}$ to be broken simultaneously for a diode response, crucially, reproducing conditions for the superconducting diode response from a many-body ground state.
\\

\textit{Microscopic origin of CANDLES: }
The diode response is quantified by the diode-efficiency
\begin{equation}
    \eta = \left|\frac{\vec J_+^c - |\vec J_-^c|}{\vec J_+^c + |\vec J_-^c|}\right|
\end{equation}
where $\vec J_{+}^c \equiv {\rm max}_{\vec q}\vec J_{\vec q}$ and $\vec J_{-}^c \equiv {\rm min}_{\vec q}\vec J_{\vec q}$ are the critical currents in two opposite directions. To obtain $\eta \neq 0$, it is crucial to break the symmetry $\vec J_{-\vec q} = -\vec J_{\vec q}$.
This symmetry is microscopically broken by the displacement of the Cooper pair center of mass which can be obtained from the modified superconducting gap inside a chiral cavity. From Eq.~(\ref{eq:GSenergy}), we obtain a change in the superconducting gap of $\delta\Delta_{\vec q} = \chi g\partial_{\Delta_\vec q}\sum_{\vec k} \langle M \rangle_{\vec k,\vec q}^{\alpha\alpha}$, and write the  self-consistent change in the superconducting gap due to chiral modes as
\begin{equation}
\label{eq:changegap}
    \delta\Delta_\vec q = \chi g\frac{U}{2}\sum_{\vec k, \alpha, \beta\neq \alpha} \frac{\langle \mathcal{M} \rangle^{\alpha\beta}_{\vec k, \vec q}\langle \check{\tau_-}\rangle^{\beta\alpha}_{\vec k,\vec q}}{E^0_{\vec k,\alpha} - E^0_{\vec k,\beta}} + {\rm c.c.}.
\end{equation}
Here $\check{\tau}_-$ is the Nambu lowering operator, see Appendix B for details. One can readily note that $\hat{\mathcal{T}}C_{\vec k} = -C_\vec k$ sets $\delta\Delta_\vec q  = 0$ in the presence of $\mathcal{I}$. However, for broken $\mathcal{I}$ and $\mathcal{T}$, $\delta\Delta_\vec q\neq 0$ leads to $\vec J_{\vec q} \neq -\vec J_{-\vec q}$.

To show the displaced Cooper pair center of mass from $\delta\Delta_{\vec q}\neq 0$, we consider a minimal Dirac model, $H_D = \hbar v_F(\zeta k_x , k_y)\cdot\vec \{\sigma_x,\sigma_y\} + \delta_{\mathcal{I}}\sigma_z$ where $v_F$ is the Fermi velocity, $\delta_\mathcal{I}\neq 0$ breaks $\mathcal{I}$, $\sigma_{x,y,z}$ is a Pauli matrix for sublattice and $\zeta = \pm$ for valleys. The operator $C_{\vec k}$ corresponding to $H_D$ evaluates to $C_D(\vec k) = 2\zeta(\hbar v_F)^2 \sigma_z$, capturing chirality-valley locked orbital magnetization which opens a valley dependent topological gap in the Dirac bands determined by the polarization of the chiral modes~\cite{wang2019cavity,masuki2023berry,dag2024engineering,jiang2024engineering}. This gives

\begin{equation}
\label{eq:modifiedgraphene}
    H_{D} \rightarrow \tilde{H}_{D} = H_{D} + \zeta \chi \delta_c  \sigma_z
\end{equation}
where $\delta_c =  2g (\hbar v_F)^2$. The valley polarization due to $\delta_c\neq 0$ is akin to a Haldane mass, and distinct from valley ferromagnertic order which has the form $\propto\zeta\sigma_0$~\cite{scammell2022theory,banerjee2024enhanced}, and requires magnetic field training. 
Assuming $\Delta_\vec q  =\Delta$, in Fig.~\ref{fig1}d,e we show the quasiparticle spectrum for $\vec k  = \{k,0\}$, $\Delta=0.2v$, $\delta_{\mathcal{I}} = 0.5 v$ and $\delta_c = 0$ (gray line), $0.15v$ ($\chi=+$ magenta line, $\chi = -$ green line). Note that particle-hole sectors in Eq.~(\ref{eq:bdgHam}) for $\tilde{H}_D$ map to $\zeta=\pm$. Without a supercurrent $(q=0)$, the spectrum is symmetric for $\delta_c=0$ and becomes asymmetric for $\delta_c\neq 0$. For $\chi = \pm$, negative (positive) energies move closer to the Fermi surface which indicates a reduction of the superconducting gap in the $\zeta=\mp$ valley. This implies that center of mass of the Cooper pairs is shifted towards $\zeta=\pm$ for $\chi=\pm$. In the presence of a supercurrent ($\vec q = \{q >0 , 0\}$), the superconducting gap for $\chi = \pm$ further decreases at $\zeta=\mp$ along (opposite to) $\vec q$. This implies that for $q>0$ the supercurrent decreases (increases) for $\chi=\pm$, underscoring the fundamental principle of CANDLES. The conclusions drawn from Fig.~\ref{fig1}c can be generalized to any superconductor, e.g., Rashba superconductors. A cavity design hosting out of plan polarization is necessary for Rashba superconductors as the in-plane polarizations do not shift the Cooper pair center of mass in these systems due to local spin-momentum locking, see Appendix C for details. \\

\textit{Chiral mode induced diode response in TBG: }
We now turn to demonstrating CANDLES in a candidate material -- TBG which hosts isolated flat Bloch bands near the Fermi energy. Given the important role of wavefunction winding in flat band superconductivity~\cite{peotta2015superfluidity,xie2020topology,verma2021optical,chen2024ginzburg,arora2024quantum}, TBG helps to understand the role of photonic modes in controlling quantum geometry and its role in diode responses. The Cooper pairs in TBG are formed by electrons with $\{K\uparrow, K' \downarrow\}$ where $K$ and $K'$ denote $\zeta=\pm$ for graphene valleys. 
We project Eq.~(\ref{eq:bdgHam}) onto the conduction flat band and, given the role of quantum geometry, the superconducting gap includes wavefunction coherence between the electrons in the two valleys, $\hat{\Delta}_{\vec q} = \Delta \Gamma_{\vec k , \vec q}$ with $\Gamma_{\vec k, \vec q} = \langle \psi_{\vec k + \vec q , K} | \psi^*_{-\vec k, K'} \rangle$ which under $\mathcal{T}$-symmetry satisfies $|\Gamma_{\vec k,\vec q}|^2 \approx 1 - q_aq_b g^{ab}_{\vec k}$; $g_{\vec k}^{ab}$ is the quantum metric (also see Appendix D for details).

We consider the continuum model for TBG which captures the low energy physics resulting from Dirac electrons in each layer. The complete description of our TBG Hamiltonian is provided in Appendix E.
Noting that minimal coupling does not affect the interlayer coupling allows us to treat the light-matter interaction qualitatively at the level of a single layer Dirac Hamiltonian~\cite{jiang2024engineering}. Thus, we use $\delta_\mathcal{I}$ and $\delta_c$ as parameters to account for $\mathcal{I},\mathcal{T}$-breaking, respectively. The hBN encapsulation of graphene layers makes $\delta_\mathcal{I}\neq 0$, and distinguished here as $\delta_{\mathcal{I}}\rightarrow\delta_{\rm hBN}^{1,2}$. 
We comment that beyond our treatment, the quantum correlations due to photonic modes can also modify the TBG band structure through inter-layer couplings, which have more considerable effects on remote bands~\cite{jiang2024engineering}. To proceed we tune $\delta_c$ in the range of hundreds of $\mu$eV to a few meV, values previously obtained using perturbation and exact diagonalization techniques~\cite{wang2019cavity,masuki2023berry,dag2024engineering}. For our system, we obtain $\delta_c = 0.6$ meV at vacuum, see Appendix F for details. We comment that the cavity induced gap is obtained corresponding to favorable parameters regime with relatively small effective mass valid for graphene and related systems. In more conventional superconductors, the resulting $\delta_c$ may be smaller. Nevertheless, the mechanism itself is general and should remain observable provided sufficiently strong light–matter coupling or suitably engineered electronic structures are realized.

\begin{figure}
    \centering
    \includegraphics[width=1\linewidth]{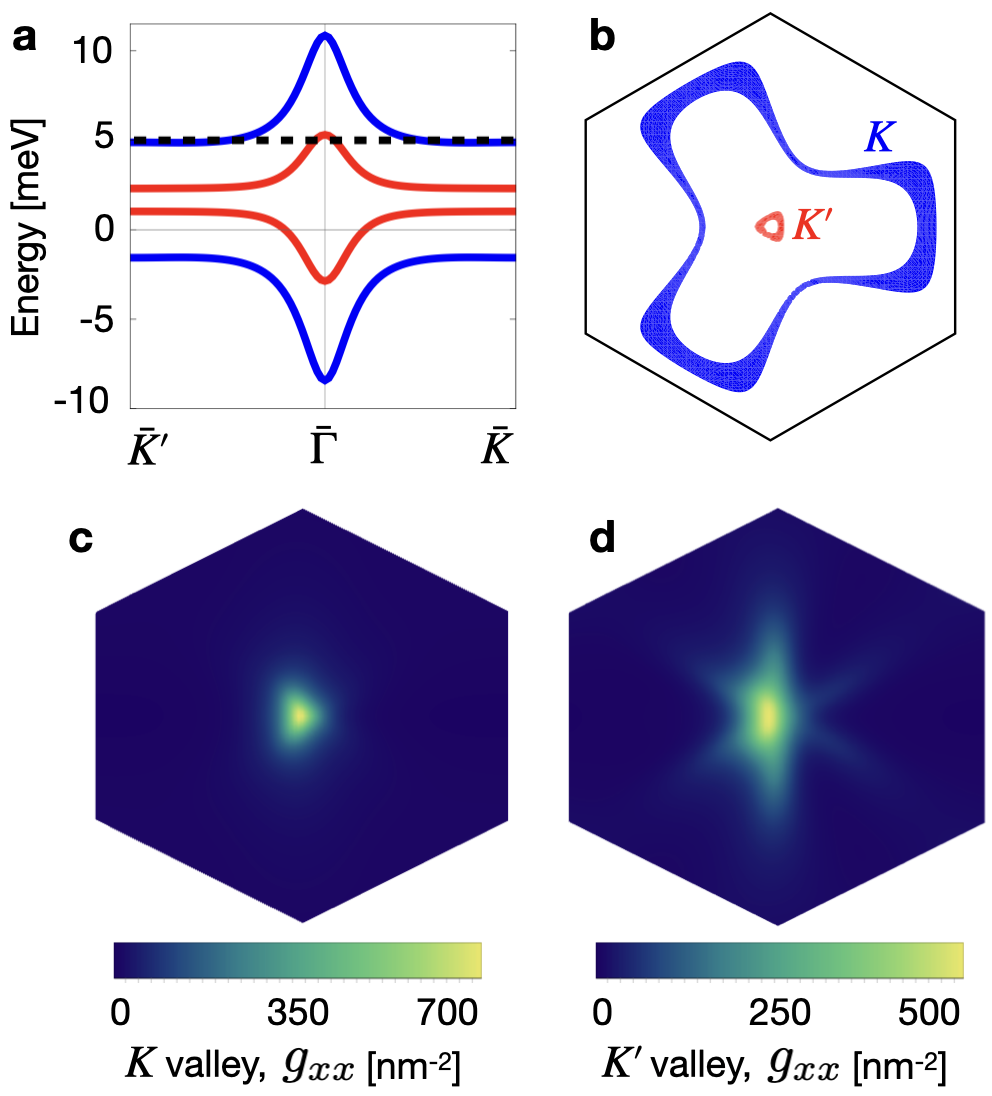}
    \caption{Time reversal symmetry breaking in hBN encapsulated TBG interacts with chiral cQED modes. (a) TBG band structure for flat bands in $K$ (blue) and $K'$ (red) valleys. Here we have used $\delta_{\rm hBN}^1 = \delta_{\rm hBN}^2 = 5$ meV, $\chi=+$, $\delta_c = 7.5$ meV. (b) Fermi surface contour ($\mu=5$ meV) in $K$ and $K'$ valleys showing valley polarization due to broken $\mathcal{T}$. (c) Quantum metric, $g^{xx}_{\vec k}$, distributions in $K$ and $K'$ valleys which give rise to a natural dipole.}
    \label{fig2}
\end{figure}

\begin{figure*}
\centerline{\includegraphics[width=\textwidth]{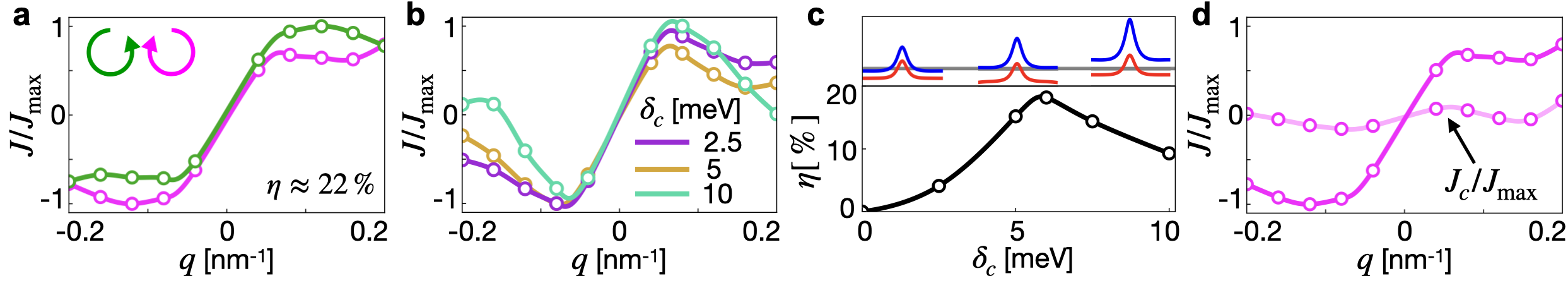}}
\caption{CANDLES in a TBG heterostructure. (a) Chirality dependent supercurrent nonreciprocity in TBG where the $J/J_{\rm max}$ for right handed modes ($\chi=+$, magenta line) has a larger supercurrent for $-q$, and for left handed modes ($\chi=-$, green line) the maximum supercurrent is obtained for $+q$. The maximum asymmetry in magnitude of $J/J_{\rm max}$ is $64\%$ which corresponds to diode efficiency of $\eta \approx 22\%$. Parameters used: $\delta_{\rm hBN}^1 =\delta_{\rm hBN}^2 = 5$ meV, $\delta_c = 5$ meV and $\mu = 2$ meV. (b) $J/J_{\rm max}$ at different values of $\delta_c$ with $\chi=+$, $\delta_{\rm hBN}^1 = \delta_{\rm hBN}^2 = 5$ meV and $\mu = 4$ meV. (c) Non-monotonic behavior of $\eta$ in variation with $\delta_c$ which can be understood from the evolution of flat bands across the Fermi level (top panel). (d) Comparison of supercurrent without quantum geometry, $J_c/J_{\rm max}$, with total supercurrent, $J/J_{\rm max}$ indicating that CANDLES in TBG is dominated by quantum geometric nonreciprocities. Parameters used, same as panel (a).  
}
\label{fig3}
\end{figure*}  
 
In Fig.~\ref{fig2}a, we show the band structure for TBG flat bands for $\delta_c = 7.5$ meV. $\mathcal{T}$-breaking by chiral modes polarizes the valleys, see Fig.~\ref{fig2}b, and skews the quantum metric distributions forming a natural intervalley dipole, see Fig.~\ref{fig2}c,d. While valley ferromagnetic states ($\propto \zeta\sigma_0$) in TBG can arise due to Coulomb interaction-driven Stoner instabilities, they require magnetic field training to stabilize~\cite{de2021gate,diez2023symmetry,lin2022zero,scammell2022theory,hu2024quantum}. Such states are limited by material retentivity and polarity reversal necessitates scanning the hysteresis loop. In contrast, a chiral mode-generated magnetic state is easily tunable by cavity parameters, and switchable on ultrafast ns-$\mu$s time scales when compared to a typical hysteresis lag of a few seconds.

The valley polarized states and the corresponding quantum geometry are central to superconducting nonreciprocity. 
We numerically calculate the supercurrent from Eq.~(\ref{eq:bdgHam}). We fix $\vec q = \{q,0\}$ without loss of generality, see Appendix G for details. In Fig.~\ref{fig3}a, we show the chirality dependence of nonreciprocal supercurrent, $J/J_{\rm max}$ where $J_{\rm max} = \rm{max}(|J|)$. Large asymmetry emerges, $J_-/J_+\approx 64\%$, for chiral light-matter hybrid states which correspond to a diode-efficieny of $\eta\approx 22\%$. Importantly, the direction of nonreciprocity is controlled by the chirality of modes: right handed modes ($\chi=+$) host a larger supercurrent for $-q$, and left handed modes ($\chi=-$) have a larger supercurrent for $+q$, consistent with Fig.~\ref{fig1}d. 

To understand the impact of cavity design, in fig.~\ref{fig3}b we show $J/J_{\rm max}$ for $\chi=+$ at different values of $\delta_c$ and sublattice potential at $\delta_{\rm hBN}^1=\delta_{\rm hBN}^2 = 5$ meV. We note a non-monotonic variation of supercurrent asymmetry in variation with $\delta_c$. The supercurrent asymmetry increases at first as we increase $\delta_c$ from 2.5 meV (purple line) to 5 meV (yellow line) but falls as we further increase $\delta_c$, e.g., $\delta_c = 10$ meV (teal line). This non-monotonic behavior can be vividly seen in fig.~\ref{fig3}c where we show $\eta$ in variation with $\delta_c$ which can be understood from the evolution of flat bands across the Fermi level with change in $\delta_c$, see top panel of Fig.~\ref{fig3}c. For small $\delta_c$ ($\delta_c= 2.5$ meV) the valleys are partially polarized, and upon increasing $\delta_c^R$ the polarization between valleys increases, ultimately leading to fully polarized valley states, e.g., $\delta_c=5$ meV which explains the increase in $\eta$. However, as $\delta_c$ is increased further, fully polarized valleys lead to a weakening of superconducting order, and the Fermi level moves away from the van Hove singularities around the $\bar{\Gamma}$-point where band curvature and the quantum metric are maximum, e.g., see Fig~\ref{fig2} which accounts for the reduced $\eta$ at higher $\delta_c$.  

Given the important role of quantum geometry in flat band superconductivity, we weigh in on the quantum geometric contribution for CANDLES in TBG flat bands. In fig.~\ref{fig3}d we show the supercurrent $J_{c}/J_{\rm max}$ without quantum geometry, i.e.,  $\Gamma(\vec k, \vec q)=1$. We compare it with the total supercurrent, $J/J_{\rm max}$; $J_{\rm max}$ is taken with respect to the total current, $J$. We note that $J_c$ barely contributes to the total supercurrent indicating that superconducting nonreciprocities shown here are dominated by quantum geometry. This is consistent with a recently proposed quantum geometric origin of the superconducting diode effect in flat band superconductors~\cite{hu2024quantum}. We also comment that cavity control of the quantum metric is seldom discussed in literature when compared to Berry curvature. The reason partly lies in the fact that Berry curvature can be directly probed in transport measurements such as anomalous Hall effect~\cite{mciver2020light}. However, measuring superconducting nonreciprocities in flat band systems provides an direct probe of the photon controlled quantum metric. 
\\

\textit{Experimental feasibility of CANDLES: }
Importantly, CANDLES is driven by virtual photon exchange, allowing the use of cavities with a wide range of frequencies. Focusing on microwave cavities, the chiral modes considered in this work and shown in Fig.~\ref{fig1}a have been realized in Ref.~\cite{owens2022chiral} where long lived chiral modes are spectrally separated and can be individually excited using gaussian pulses. A similar configuration of chiral modes has been realized in Ref.~\cite{rosen2024synthetic} where chiral modes are realized through parametric coupling on a 4x4 quantum processor.  

Given the realizable cavity platforms, we propose a method to quantify CANDLES using inductive measurements in lumped circuits which are fundamental for microwave devices, and enable extraction of nonreciprocal superfluid stiffness. In a recent experiment~\cite{tanaka2024kinetic}, it was found that for an applied current, $I_a$ the TBG inductance behaved as $L \approx L_0[1 + (I_a/I_{\rm max})^2]$. The corresponding frequency measured by a lumped element is $f=f_0\sqrt{L_0/L}$ where $f_0 = 1/\sqrt{L_0C}$ is the characteristic frequency of the circuit. The shift in resonance is $\Delta f/f_0\approx -0.5(I_a/I_{\rm max})^2$. For a diode response, we have $I_+ \neq I_-$ leading to $L_+\neq L_-$. For $\eta\approx 22\%$, we have $I_+/I_- = 5/3$ which corresponds to $L_K^- - L_K^+ = 16L_0I_a^2/(25I_-^2)$. We define the resolution frequency as $f_{\rm reso} = \Delta f_+ - \Delta f_-$. We use the values from the same experiment~\cite{tanaka2024kinetic}, $f_0=5$ GHz, $I_a = 40$ nA and $I_- = 160$ nA which corresponds to $f_{\rm reso} \approx 0.1$ GHz which is well within the range of state-of-the-art devices. Particularly, it was found that for a TBG integrated lumped circuit, the quality factor was around 1000~\cite{tanaka2024kinetic} which gives us a minimum resolution bound of 5 MHz for existing devices. Our estimates are well above the bound set by the device quality factor, underscoring its relevance to current setups~\cite{tanaka2024kinetic,banerjee2024superfluid,kreidel2024measuring}. 

\textit{Conclusion: }
This work establishes how chiral photon exchange modifies the superconducting ground-state energy, providing a universal principle for photon-tunable superconducting nonreciprocity, valid across different platforms including TBG, Rashba superconductors, and aluminum devices with conformal-mapped nanoholes~\cite{lyu2021superconducting}. CANDLES provides a magnet-free, dynamically switchable diode suitable for scalable integration in quantum circuits, and can be harnessed to supplement existing quantum technology, e.g., for efficient quantum batteries~\cite{ahmadi2024nonreciprocal}, manipulation of photonic gates~\cite{krastanov2021room}, and passive error correction~\cite{rymarz2021hardware}. Additionally, given the sizable diode efficiency, CANDLES can potentially be used as a sensitive probe for $\mathcal{T}$-breaking in hybrid circuit quantum electrodynamic devices. Importantly, the proposed functionality holds the potential to produce high quality nonreciprocal waveguides, isolators, circulators, and metasurfaces, reducing their cryogenic footprints.

\textit{Acknowledgments: }
The authors acknowledge useful discussions with William Oliver, Joel Wang, Alex Ruichao Ma, Justin Song, Emily Been, \addAA{Vasil Rokaj, Ceren Dag} and Ioannis Petrides. This work was supported by the Defense Advanced Research Projects Agency (DARPA). We also acknowledge Grant Numbers GBMF8048 and GBMF12976 from the Gordon and Betty Moore Foundation and support from the John Simon Guggenheim Memorial Foundation (Guggenheim Fellowship).

\appendix

\subsection*{Appendix A: Perturbation analysis of $\mathcal{T}$-breaking: derivation of Eq.~(\ref{eq:intHamiltonian})}
{\color{black}
Here, we summarize the key points of the analysis for $\mathcal{T}$-breaking due to chiral modes in Eq.~\ref{eq:intHamiltonian}. The light-matter coupling is accounted for via $\vec k \rightarrow \vec k-e\hat{\vec A}/\hbar$. Here, $\hat{\vec A} = \sum_{\lambda}\vec A_0^\lambda(\vec e_{\lambda} a + \vec e_\lambda^*a^\dag)$ is the vector field for modes with chirality $\vec e_{\lambda \in\{R,L\}} = (1,\chi i)/\sqrt{2}$ where $\chi=\pm$ for right ($R$) and left ($L$) handed modes, respectively. We assume complete chirality of the modes, i.e., the amplitude for one of the modes is set to zero. The total Hamiltonian of the light-matter hybrid system can then be written as 
\begin{equation}
    H_{\rm total} = H^0_\vec k +  \hbar\omega_c a^\dagger a + H_{\rm int}.
\end{equation}
where the Bloch Hamiltonian satisfies $H_{\vec k}|u_p^{\vec k}\rangle = \epsilon_p^{\vec k} |u_p^{\vec k}\rangle$ for Bloch band $p$. We use standard perturbation theory to expand the interaction Hamiltonian order by order in terms of $eA_0/\hbar$. To leading order, the interaction Hamiltonian $H_{\rm int}$ is
\begin{equation}
    H_{\rm int}^{(1)} = -\frac{e A_0}{\hbar} (\vec e \cdot \hat{\vec v} a + \vec e^* \cdot \hat{\vec v} a^\dagger) = H_{\rm int}^- a + H_{\rm int} a^\dagger
\end{equation}
where $\hat{\vec v} = \partial_{\vec k}H_{\vec k}^0$. Next, we project the interaction Hamiltonian onto the Fock state satisfying $\langle a\rangle_{\rm Fock} = \langle a^\dag \rangle_{\rm Fock} = 0$ and $\langle a^\dag a \rangle_{\rm Fock} = n$ where $n$ is the photon population where $\langle \mathcal{O}\rangle_{\rm Fock} = \langle n |\mathcal{O} | n \rangle$. Thus, $H_{\rm int}^{(1)}$ identically vanishes. We can account for the next leading order term in the interaction Hamiltonian 
\begin{equation}
\label{eq:hint2}
    H_{\rm int}^{(2)} = \sum_{q\neq p} \left[n\frac{H_{\rm int}^- |u_q^{\vec k}\rangle\langle u_q^{\vec k}| H_{\rm int}^+}{\epsilon_{q}^{\vec k} - \epsilon_{p}^{\vec k} - \hbar\omega_c} + (n+1)\frac{H_{\rm int}^+ |u_q^{\vec k}\rangle\langle u_q^{\vec k}| H_{\rm int}^-}{\epsilon_{q}^{\vec k} - \epsilon_{p}^{\vec k} + \hbar\omega_c}\right].
\end{equation}
Moving on, we split the numerator of Eq.~(\ref{eq:hint2}) into symmetric and anti-symmetric parts
\begin{equation}
    H^{\mp}_{\rm int}|u_q^{\vec k}\rangle\langle u_q^{\vec k}| H_{\rm int}^\pm = \frac{1}{2} \{H_{\rm int}^{\mp},H_{\rm int}^{\pm}\}_q + \frac{1}{2} [H_{\rm int}^{\mp},H_{\rm int}^{\pm}]_q
\end{equation}
where 
\begin{equation}
    \{H_{\rm int}^{\mp},H_{\rm int}^{\pm}\}_q = H^{\mp}_{\rm int}|u_q^{\vec k}\rangle\langle u_q^{\vec k}| H_{\rm int}^\pm + H^{\pm}_{\rm int}|u_q^{\vec k}\rangle\langle u_q^{\vec k}| H_{\rm int}^\mp 
\end{equation}
and
\begin{equation}
    [H_{\rm int}^{\mp},H_{\rm int}^{\pm}]_q = H^{\mp}_{\rm int}|u_q^{\vec k}\rangle\langle u_q^{\vec k}| H_{\rm int}^\pm - H^{\pm}_{\rm int}|u_q^{\vec k}\rangle\langle u_q^{\vec k}| H_{\rm int}^\mp.
\end{equation}
Note that $\{H_{\rm int}^{\mp},H_{\rm int}^{\pm}\}_q$ is even in $\mathcal{T}$ and only produces a rigid shift in energy. Thus, we drop it for rest of our analysis. On the contrary, $[H_{\rm int}^{\mp},H_{\rm int}^{\pm}]_q$ is odd in $\mathcal{T}$, and will produce a non-trivial effect. We utilize the fact that $\langle u_p^{\vec k} |[H_{\rm int}^{\mp},H_{\rm int}^{\pm}]_p |u_p^{\vec k}\rangle =0$, and at finite $\omega_c$, Eq.~(\ref{eq:hint2}) vanishes for $p=q$. This allows us to include the $p=q$ contribution in the summation over $q$ in Eq.~(\ref{eq:hint2}). For brevity, we set $\epsilon_q^{\vec k} - \epsilon_p^{\vec k} = \Xi$ as the energy scale determined by the dominant contribution set by the lowest possible virtual transition. Next, we work in the limit of $\hbar\omega_c \ll \Xi$ which is a valid approximation as we work at microwave frequencies. This helps us expand the denominator in Eq.~(\ref{eq:hint2}) as 
\begin{equation}
    \frac{1}{\Xi \pm \hbar\omega_c} \approx \frac{1}{\Xi} \mp \frac{\hbar\omega_c}{\Xi^2},
\end{equation}
and plugging it back into Eq.~(\ref{eq:hint2}) we get 
\begin{equation}
    H_{\rm int}^{(2)} = -  
    \frac{\hbar\omega_c}{2}\frac{[H_{\rm int}^{\mp},H_{\rm int}^{\pm}]}{\Xi^2}, \quad {\rm for}\quad \Xi\neq 0
\end{equation}
which gives the leading order non-trivial correction due to broken $\mathcal{T}$-symmetry. We simplify $H_{\rm int}^{(2)}$ further by substituting in for $H_{\rm int}^\pm$ and get 
\begin{equation}
    H_{\vec k}^1 = \chi\frac{e^2 A_0^2}{2\hbar^2}(1+2n) \hbar\omega_c \frac{i[v_x,v_y]}{\Xi^2} = \chi g(A_0,n) C_{\vec k}
\end{equation}
reproducing Eq.~(\ref{eq:intHamiltonian}).
The effective Hamiltonian, up to leading order, can then be written as 
\begin{equation}
    H_\vec k = H_\vec k^0 + H^1_\vec k.
\end{equation}
}

\subsection*{Appendix B: Chiral mode-modified superconducting ground state energy and superconducting gap}
We start from the Nambu basis Hamiltonian in Eq.~(\ref{eq:bdgHam}) for a superconducting phase which satisfies $\mathcal{H}_{\rm BdG}|\Psi_{\vec k, \vec q,\alpha}\rangle = E_{\vec k, \vec q, \alpha}|\Psi_{\vec k, \vec q,\alpha}\rangle$. The superconducting ground state is $E_{\rm GS}^\vec q = -\frac{1}{2}\sum_{\vec k,\alpha}E_{\vec k,\vec q,\alpha} + |\Delta_\vec q|^2/\mathcal{U}$ for pairing potential $\mathcal{U}$. The corresponding mean-field Hamiltonian without the cavity is $\mathcal{H}_{\rm BdG}^0$ with $\mathcal{H}_{\rm BdG}^0|\Psi_{\vec k, \vec q,\alpha}^0\rangle = E_{\vec k, \vec q, \alpha}^0|\Psi_{\vec k, \vec q,\alpha}^0\rangle$ and ground state energy $E_{\rm GS,0}^\vec q = -\frac{1}{2}\sum_{\vec k,\alpha}E_{\vec k,\vec q,\alpha}^0 + |\Delta_\vec q|^2/\mathcal{U}$ for pairing potential $\mathcal{U}$. The cavity contribution in the Nambu basis can then be separated as $\mathcal{H}_{\rm BdG}^1 = \mathcal{H}_{\rm BdG} - \mathcal{H}_{\rm BdG}^0$ where explicitly we write
\begin{equation}
    \mathcal{H}_{\rm BdG}^1 = \chi g\begin{pmatrix}
        C_{\vec k + \vec q/2} & 0 \\
        0 & -C_{-\vec k + \vec q/2}^T
    \end{pmatrix}= \chi g \mathcal{M}.
\end{equation}
Since, we assume weak coupling in the main text, the changes to the ground state energy are obtained by Taylor expanding $E_{\rm GS}$ to leading order in $g$. To explicitly write down the form of the chiral mode contribution, we use $E_{\vec k, \vec q, \alpha} = \langle \Psi_{\vec k, \vec q, \alpha}|H_{\rm BdG}|\Psi_{\vec k, \vec q, \alpha}\rangle$ which then gives 
\begin{equation}
    E_{\rm GS}\approx E_{\rm GS,0} -\frac{1}{2} \sum_{\vec k,\alpha}\chi g\langle\mathcal{M}\rangle_{\vec k,\vec q}^{\alpha\alpha}
\end{equation}
where the matrix element $\langle\mathcal{M}\rangle_{\vec k,\vec q}^{\alpha\beta}$ is given in Eq.~(\ref{eq:GSenergy}) of the main text. 

Next, we calculate the self-consistent change in the superconducting gap due to the chiral modes shown in Eq.~(\ref{eq:changegap}) of the main text. We begin with the variational condition to obtain $\Delta$
\begin{equation}
    \partial_{\Delta^*_\vec q}E_{\rm GS} = 0
\end{equation}
which gives
\begin{equation}
    \Delta_\vec q = \frac{\mathcal{U}}{2}\sum_{\vec k,\alpha}\partial_{\Delta^*_\vec q}E_{\rm GS}
\end{equation}
This is exact. We again use Hellman-Feynman theorem to write
\begin{equation}
    \partial_{\Delta^*}E_{\vec k, \vec q, \alpha} = \langle\Psi_{\vec k, \vec q, \alpha}| \partial_{\Delta^*_\vec q} \mathcal{H}_{\rm BdG}|\Psi_{\vec k, \vec q,\alpha}\rangle = \begin{pmatrix}
        0 & 0 \\
        \check{\tau}_- & 0 
    \end{pmatrix}
\end{equation}
where $\check{\tau}_- = \tau_x - i\tau_y$ is the Nambu lowering operator; $\tau_{x,y,z}$ is a Pauli matrix in the Nambu basis. This reproduces the familiar self-consistent gap equation at zero temperature
\begin{equation}
\label{eq:selfgap}
    \Delta_{\vec q} = \frac{\mathcal{U}}{2}\sum_{\vec k,\alpha} \langle\Psi_{\vec k, \vec q,\alpha}|\check{\tau}_-|\Psi_{\vec k, \vec q,\alpha}\rangle.
\end{equation}
Moving forward, we now expand the Nambu state to leading order to $g$
\begin{equation}
   | \Psi_{\vec k, \vec q,\alpha}\rangle \approx | \Psi_{\vec k, \vec q,\alpha}^0\rangle + \chi g\sum_{\beta\neq\alpha}\frac{\langle \mathcal{M}\rangle_{\vec k, \vec q}^{\alpha\beta}}{E_{\vec k, \vec q, \alpha}^0 - E_{\vec k, \vec q, \alpha}^0}
\end{equation}
which when plugged into Eq.~(\ref{eq:selfgap}) reproduces Eq.~(\ref{eq:changegap}) in the main text. 

\subsection*{Appendix C: Conditions for CANDLES in Rashba superconductors}
Rashba superconductors were first the class of materials to exhibit the superconducting diode effect. A simple model of a 2D electron gas with Rashba spin orbit coupling (SOC) is
\begin{equation}
\label{eq:rashbaHam}
    H_{R} = \frac{\hbar^2 |\vec k|^2}{2m^*} + \alpha_R(s_x k_y - s_y k_x)
\end{equation}
where $m^*$ is the effective mass of the electron gas and $\alpha_R$ is the parameter for Rashba SOC. Using Eq.~(\ref{eq:intHamiltonian}) directly gives $C_R = 2\alpha_R^2 s_z$ and produces $\mathcal{T}$-breaking similar to Zeeman splitting due to an out-of-plane magnetic field. However, it is known that such a Zeeman splitting cannot produce a diode response due to the lack of a Lifshitz invariant order parameter gradient constraining $\vec J_\vec q = \vec J_{-\vec q}$~\cite{yuan2022supercurrent}. However, the effects of chiral modes derived in the main text can be readily generalized by reconsidering the cavity design such that an out-of-plane polarization is sustained. From an experimental point of view, this can be readily done at microwave-frequencies with 3D cylindrical cavities, dielectric resonators, and split-ring resonators. 

We consider chiral modes with $A_x$ and $A_z$ with a phase difference of $\pi/2$. Then, we calculate $\tilde{C}_R = \chi g \alpha_R^2 s_y$ which produces a Zeeman splitting analogous to a magnetic field along the $y$-direction, generating a diode response along the $x$-direction~\cite{yuan2022supercurrent}. We modify the Hamiltonian in Eq.~(\ref{eq:rashbaHam}) as
\begin{equation}
    \tilde{H}_R \rightarrow H_R + \delta_{c}^R s_y
\end{equation}
where $\delta_c^R$ is the parameter capturing the effects of chiral modes. Beyond this point, Eq.~(\ref{eq:bdgHam}-\ref{eq:changegap}) can be generically employed to evaluate CANDLES in Rashba superconductors. 

{\color{black}
\subsection*{Appendix D: Description of flat band superconductivity}
Here we describe the fundamentals of flat band superconductivity. We consider a TBG heterostructure as a multiband system which hosts isolated flat Bloch bands near the Fermi energy, $\mu$. Assuming that the separation of a single flat band to other bands is greater than its bandwidth, we map the system onto the TBG flat band. The Hamiltonian near the Fermi energy can then be written as $H_{0} = \xi_{\vec k,s} c_{\vec k,s}^\dag c_{\vec k,s}$ where $c_{\vec k,s}^\dag$ ($c_{\vec k,s}$) creates (annihilates) electrons only in the flat band, and $\xi_{\vec k} = \epsilon_{\vec k} - \mu$ with $\epsilon_{\vec k}$ being the energy of Bloch state $|\psi_{\vec k}\rangle$ obtained by solving $H_0|\psi_{\vec k}\rangle = \epsilon_{\vec k}|\psi_{\vec k}\rangle$. Here, $s$ denotes time-reversal pairs participating in an attractive pairing channel. For TBG $s\in\{K\uparrow, K' \downarrow\}$ where $K$ and $K'$ denote two graphene valleys. The effective attractive interaction is 
\begin{equation}
    H_{\rm int} = \sum_{\vec k} \Gamma(\vec k, \vec q) \Delta_{\vec q} c_{\vec k + \vec q, s}^\dagger c_{\vec k, \bar{s}} + {\rm h.c.} + |\Delta_{\vec q}|^2/\mathcal{U}
\end{equation}
where $\mathcal{U}$ is the strength of the interaction mediating superconductivity, $\Delta_{\vec q}$ is the order parameter and $\vec q$ is the Cooper pair momentum denoting a current carrying state. Importantly, 
\begin{equation}
    \Gamma(\vec k, \vec q) = \langle \psi_{\vec k + \vec q , s} | \psi^*_{-\vec k, \bar{s}} \rangle
\end{equation}
is the quantum distance between time reversal pairs $s$ and $\bar{s}$, 
encoding the geometry of Bloch states, known to play a crucial role in  flat band superconductivity~\cite{peotta2015superfluidity,xie2020topology,verma2021optical,chen2024ginzburg,arora2024quantum}.
The wavefunction coherence $\Gamma(\vec k,\vec q)$ under $\mathcal{T}$-symmetry satisfies $|\Gamma(\vec k,\vec q)|^2 = 1-q_aq_b g^{ab}_{\vec k}$ where $g_{\vec k}^{ab}$ is the quantum metric.
}

\subsection*{Appendix E: Continuum model for the TBG-hBN heterostructure}
We follow Ref.~\cite{koshino2018maximally} to model TBG. In each graphene layer, the primitive (original) lattice vectors are $\vec{a}_1 = a_{\rm G}(1,0)$ and $\vec{a}_2 = a_{\rm G}(1/2,\sqrt{3}/2)$ with $a_{\rm G}=0.246$ nm being the lattice constant. The corresponding reciprocal space lattice vectors are $\vec{b}_1 = (2\pi/a_{\rm G})(1,-1/\sqrt{3})$ and $\vec{b}_2 = (2\pi/a_{\rm G})(0,2/\sqrt{3})$, and Dirac points are located at $K_\zeta = -\zeta(2\vec{b}_1 +\vec{b}_2)/3$. For a twist angle $\theta$ (accounting for the rotation of layers), the lattice vectors of layer $l$ are given by $\vec{a}_{l,i} = R(\mp\theta/2)\vec{a}_i$, $\mp$ for $l=1,2$ respectively, and $R(\theta)$ represents a rotation by an angle $\theta$ about the normal. Also, from $\vec{a}_{l,i}.\vec{b}_{l',j} = 2\pi \delta_{ij}\delta_{ll'}$ we can check that the reciprocal lattice vectors become $\vec{b}_{l,i} = R(\mp\theta/2)\vec{b}_i$ with corresponding Dirac points now located at $\vec{K}_{l,\zeta} =  -\zeta(2\vec{b}_{l,1} +\vec{b}_{l,2})/3$.

At small angles, the slight mismatch of the lattice period between two layers gives rise to long range moir\'e superlattices. The reciprocal lattice vectors for these moir\'e superlattices are given as $\vec{g}_i = \vec{b}_{1,i} - \vec{b}_{2,i}$. The superlattice vectors $\vec{L}$, can then be found using $\vec{g}_i.\vec{L}_j = 2\pi \delta_{ij}$, where $\vec{L}_1$ and $\vec{L}_2$ span the moir\'e unit cell with lattice constant $L = \vec{L}_1 = \vec{L}_2 = a_{\rm G}/[2\sin\theta/2]$. 

Next, when the moir\'e superlattice constant is much longer than the atomic scale, the electronic structure can be described using an effective continuum model for each valley $\zeta=\pm$. The total Hamiltonian is block diagonal in the valley index, and for each valley, the effective Hamiltonian in the continuum model is written in terms of the sublattice and layer basis $(A_1, B_1, A_2, B_2)$
\begin{equation}
\label{eq:tbgHam}
H_\zeta = 	\begin{pmatrix}
H_{1,\zeta}(\vec{p}) & T^\dag_\zeta \\
T_\zeta & H_{2,\zeta}(\vec{p})
\end{pmatrix}
\end{equation}
where $H_{l,\zeta} = -\hbar v_F R(\pm \theta/2)\vec{p}.(\zeta\sigma_x, \sigma_y)$ is the Hamiltonian for each layer with $\hbar v_F/a_{\rm G} = 2135.4$ meV, and 
\begin{multline}
T_\zeta = \begin{pmatrix}
t_{\rm AA} & t'_{\rm AB} \\
t'_{\rm AB} & t_{\rm AA} 
\end{pmatrix} +
\begin{pmatrix}
t_{\rm AA} & t'_{\rm AB}e^{-i\zeta\frac{2\pi}{3}} \\
t'_{\rm AB}e^{i\zeta\frac{2\pi}{3}} & t_{\rm AA} 
\end{pmatrix}e^{i\zeta \vec{g}_1.\vec{r}}\\
+ 
\begin{pmatrix}
t_{\rm AA} & t'_{\rm AB}e^{i\zeta\frac{2\pi}{3}} \\
t'_{\rm AB}e^{-i\zeta\frac{2\pi}{3}} & t_{\rm AA}
\end{pmatrix}e^{i\zeta (\vec{g}_1+\vec{g}_2).\vec{r}}
\end{multline}
where $\vec{g}_i$ is the reciprocal lattice vector of the mBZ. In what follows, we use the tunneling parameters $t_{\rm AB}=79.7$ meV and $t'_{\rm AB}=97.5$ meV. When hBN is aligned with the graphene layers, $C_2$ symmetry is broken, modifying the layer Hamiltonians $H_{l,\zeta}$. This can be described by introducing a sublattice staggered potential $\delta_l$ so that the Hamiltonian for each layer $H_{l,\zeta}(\vec{p}) \rightarrow H_{l,\zeta}(\vec{p})+\delta_l\sigma_z$.

{\color{black}
\subsection*{Appendix F: Estimates for $\delta_c$}
In the main text, we defined the chiral cavity gap as  $\delta_c = 2g(\hbar v_F)^2$. However, as shown in Ref.~\cite{dag2024engineering}, to correctly estimate the order of light-matter coupling strength for the Dirac Hamiltonian, an input parameter determined by experiments is required to account for differences between the free electron mass and the effective mass of Dirac electrons. This estimation of the order of light-matter coupling was verified in Ref.~\cite{tay2025terahertz} using matrix elements calculated from density functional theory (DFT). We include the input parameters such that 
\begin{equation}
    g \rightarrow \tilde{g} = (1+2n)\left(\frac{eA_0}{\hbar m \Xi}\right)^2 \hbar\omega_c
\end{equation}
where $m$ is the dimensionless input parameter. Here we use, $m=0.01$ within the range of values considered in Ref.~\cite{dag2024engineering}. We plug in the numbers to obtain the gap opened at vacuum ($n=0$)
\begin{align}
    \delta_c &= \frac{1}{2} \left(\frac{e^2}{\epsilon_0}\right) \left(\frac{\hbar v_F}{\Xi}\right)^2 \frac{1}{V} \nonumber \\
    & = \frac{1}{2} (1.8 \times 10^{-5} \text{ meV m})\frac{(6.6\times 10^{-7} \text{ meV m})^2}{(1 \text{ meV})^2}\frac{1}{10^{-14}\text{ m}^3} \nonumber \\
    & = 0.6 \text{ meV}.
\end{align}
Other parameters used in the above calculation are $v_F = 10^6 \text{ m/s}$ and $\Xi = 1$ meV. We have used a mode confinement of $V=10^{-14}\text{ m}^3$ which is typical for co-planar microwave cavities~\cite{blais2004cavity}. 
}

\subsection*{Appendix G: Numerical evaluation of diode response}
The band structure and quantum metric for TBG were obtained numerically by diagonalizing the continuum Hamiltonian, details for which are presented in SM section S5. The continuum Hamiltonian is truncated such that at each $\vec k$ in the moir\'e Brillouin Zone (mBZ), a $244\times 244$ Hamiltonian is diagonalized ensuring convergence of the eigenvalue problem for flat bands.

The supercurrent for TBG is calculated using the ground state corresponding to Eq.~(\ref{eq:bdgHam}) such that $J_{\vec q} = \partial_{\vec q} E_{\rm GS}$. The corresponding Bloch eigenvalues and eigenstates are calculated using Eq.~(\ref{eq:tbgHam}), supplemented with $\mathcal{T}$-breaking in Eq.~(\ref{eq:modifiedgraphene}). We fix the original superconducting gap at 1 meV, $T=0$, $\vec q = \{q,0\}$. At each value of $q$, we numerically differentiate $E_{\rm GS}$ and evaluate the Riemann sum over $\vec k$ on a discrete grid in the $(k_x, k_y)$ plane of 2500 points in the mBZ.  
\bibliography{ref}

\end{document}